\documentclass{ws-ijmpd}
\usepackage{tablefootnote}
\usepackage[super,compress]{cite}
\begin{document}

\markboth{Alessandro Gruppuso, Sperello di Serego Alighieri}
{Cosmic Polarisation Rotation from CMB data}

%
\catchline{}{}{}{}{}
%

\title{COSMIC POLARISATION ROTATION\footnote{Note that in the literature, the term ``Cosmic Birefringence'' is commonly used to refer to this phenomenon.} 
$~$FROM CMB DATA: \\ A REVIEW FOR GR110\footnote{This paper is to be published also in the book “One Hundred and Ten Years of General Relativity: From Genesis and Empirical Foundations to Gravitational Waves, Cosmology and Quantum Gravity,”edited by Wei-Tou Ni (World Scientific, Singapore, 2025)}}

\author{ALESSANDRO GRUPPUSO\footnote{Corresponding author.}}
\address{Istituto Nazionale di Astrofisica, Osservatorio di Astrofisica e Scienza dello Spazio di Bologna, \\ via Piero Gobetti 101, 
40129, Bologna
Italy, \\ alessandro.gruppuso@inaf.it}
\address{Istituto Nazionale di Fisica Nucleare, Sezione di Bologna, \\ Viale Berti Pichat 6/2, I-40127 Bologna, Italy}
\address{Dipartimento di Fisica e Scienze della Terra, Università degli Studi di Ferrara, \\ Via G. Saragat 1, I-44122 Ferrara, Italy}

\author{SPERELLO di SEREGO ALIGHIERI}
\address{Accademia delle Scienze dell’Umbria, \\ Viale Roma 15, 06121 Perugia, Italy \\
sperello52@gmail.com}

\maketitle


\begin{abstract}
We provide an update to Ref.~\refcite{diSeregoAlighieri:2015tgl}, focusing on recent developments regarding constraints on Cosmic Polarization Rotation (CPR), also known as Cosmic Birefringence (CB), derived from Cosmic Microwave Background (CMB) polarization data. 
\end{abstract}

\keywords{Violation of parity symmetry in the electromagnetic sector; CMB polarisation.}

\ccode{PACS numbers: 98.80.$-$k, 98.70.Vc, 42.25.Ja}


\section{Introduction}	

Cosmic Polarization Rotation (CPR), also known in the literature as Cosmic Birefringence (CB), refers to the rotation of the linear polarization plane of photons in a vacuum \cite{Carroll:1989vb,Carroll:1991zs,Harari:1992ea,Carroll:1998zi,Ni:2004jj,Kostelecky:2007zz,Ni:2007ar}. 
According to standard Maxwellian theory, linearly polarized electromagnetic waves should not rotate in a vacuum, as parity symmetry is conserved. However, in extensions of electromagnetism that violate parity symmetry, such a rotation becomes possible\cite{Ni:1977zz,Ni:2016ijmpd}. 
A well-known example of this is provided by Chern-Simons theory \cite{Carroll:1989vb}.

The Cosmic Microwave Background (CMB) radiation is linearly polarized due to Thomson scattering at the last scattering surface, located at a redshift of $z \sim 1100$.
This makes the CMB an ideal probe for detecting CPR/CB, as it represents the most distant source of linear polarization observable in nature. Even tiny effects could accumulate over cosmological distances, potentially making them observable \cite{Komatsu:2022nvu}.
From an observational perspective, CPR/CB can be classified into two types: isotropic and anisotropic. In the anisotropic case, the rotation angle varies depending on the direction of observation, whereas in the isotropic case, the angle is uniform across all directions.

In this review, we provide an update on the constraints of this effect as previously reported by di Serego Alighieri in Ref.~\refcite{diSeregoAlighieri:2015tgl}, focusing on recent results concerning both isotropic and anisotropic CPR/CB derived from CMB observations.
The review is organized as follows:
In Section \ref{isotropic}, we focus on the isotropic CPR/CB effect and describe recent constraints, most of which have been obtained using a new method, the Minami-Komatsu (MK) technique \cite{Minami:2019ruj,Minami:2020fin}.
In Section \ref{Anisotropic}, we present recent constraints on the anisotropic CPR/CB effect.
Finally, in Section \ref{Conclusions}, we provide concluding remarks and discuss future CMB measurements in this context.

\section{Recent Constraints on the Isotropic CPR/CB Effect}
\label{isotropic}

The relation used in the literature to constrain the isotropic CPR/CB angle from CMB spectra is the following \cite{Lue:1998mq,Feng:2004mq,Liu:2006uh}: 
\begin{eqnarray} 
C_\ell^{TT, \text{obs}} &=& C_\ell^{TT}, \label{TTobs} \\ 
C_\ell^{TE, \text{obs}} &=& C_\ell^{TE}\cos(2\beta), \label{TEobs} \\ 
C_\ell^{TB, \text{obs}} &=& C_\ell^{TE}\sin(2\beta), \label{TBobs} \\ 
C_\ell^{EE, \text{obs}} &=& C_\ell^{EE}\cos^2(2\beta) + C_\ell^{BB}\sin^2(2\beta), \label{EEobs}\\ 
C_\ell^{BB, \text{obs}} &=& C_\ell^{BB}\cos^2(2\beta) + C_\ell^{EE}\sin^2(2\beta), \label{BBobs}\\ 
C_\ell^{EB, \text{obs}} &=& \frac{1}{2} \left(C_\ell^{EE} - C_\ell^{BB}\right)\sin(4\beta), \label{EBobs} 
\end{eqnarray} 
where $C_\ell^{X, \text{obs}}$ represents the observed CMB spectra in the presence of CPR/CB (parametrised by the angle $\beta$), and $C_\ell^{X}$ the spectra expected in the absence of CPR/CB, i.e., when $\beta=0$, with $X$ indicating the specific spectrum considered, i.e. $X=TT$\footnote{As expected, $TT$ does not depend on $\beta$ since the phenomenon under consideration affects only linear polarization.}, $TE, EE, BB, TB, EB$. Eqs. (\ref{TEobs})–(\ref{EBobs}) apply in the ideal case where the global instrumental polarization angle $\alpha=0$. However, in real CMB experiments, $\alpha$ is known only up to a given accuracy. Thus, the above equations should be modified by replacing $\beta$ with $\beta + \alpha$, as both angles affect the CMB spectra in the same way \cite{Keating:2012ge}. This introduces a perfect degeneracy between the two angles, which can only be resolved considering additional information\footnote{Note that, for example, in the Chern-Simons extension of electromagnetism (henceforth referred to as CS theory) \cite{Carroll:1989vb}, the CPR/CB effect impacts the CMB spectra as shown in Eqs.~(\ref{TEobs})–(\ref{EBobs}) if the axion-like field has a sufficiently ultralight mass. If this condition is not met, Eqs.~(\ref{TEobs})–(\ref{EBobs}) are no longer valid for $\beta$, see Refs.~\refcite{Finelli:2008jv,Fedderke:2019ajk,Galaverni:2023zhv}, but remain applicable for $\alpha$, thereby alleviating the degeneracy. See also analyses on the tomographic approach to CPR/CB, as discussed in Refs.~\refcite{Sherwin:2021vgb,Nakatsuka:2022epj,Eskilt:2023nxm}.}.

In Ref.~\refcite{Planck:2016soo}, using a harmonic estimator\cite{QUaD:2008ado,Gruppuso:2016nhj}, the Planck collaboration estimated 
$ \beta = 0.31 \pm 0.05 \, (\mbox{stat}) \pm 0.28 \, (\mbox{sys}) \, \mbox{ [deg]}, $ 
where ``sys" denotes uncertainty in the global instrumental polarization. In this case, the degeneracy was resolved by an independent estimate of $\alpha$ based on ground measurements \cite{Rosset:2010vc}. Following the approach suggested in Ref.~\refcite{Keating:2012ge}, the Polarbear \cite{POLARBEAR:2019kzz}, SPT \cite{SPT:2020cxx}, and ACT \cite{Namikawa:2020ffr} collaborations “self-calibrate” the global instrumental polarization: this method, which involves nulling the EB spectrum, effectively provides an estimate of the sum $\alpha + \beta$, as shown in Table \ref{table1}. 
Note, however, that in Ref.~\refcite{Murphy:2024fna}, it was shown that, using optical modeling, overall systematic uncertainties associated with the instrument polarization corrections for the ACTPol arrays can be provided at approximately $0.1  \, \mbox{[deg]}$.
The convention used in Table \ref{table1} follows the standard IAU convention \cite{diSeregoAlighieri:2016lbr}.


\begin{table}[ph]
\tbl{Recent constraints on the isotropic CPR/CB effect from CMB data. Note that the angles are provided according to the IAU convention.}
{\begin{tabular}{@{}cccc@{}} \toprule
angle & estimate [deg] & data with reference & method  \\
 \colrule
$\beta + \alpha$ & $\hphantom{+}0.61 \pm 0.22\hphantom{0}$  & Polarbear, Ref.\cite{POLARBEAR:2019kzz} & EB-nulling \\
$\beta + \alpha$ & $-0.63 \pm 0.04\hphantom{0}$  & SPT, Ref.\cite{SPT:2020cxx} & EB-nulling \\
$\beta + \alpha$ & $-0.12 \pm 0.06\hphantom{0}$  & ACT, Ref.\cite{Namikawa:2020ffr} & EB-nulling \\
$\beta$ & $-0.07 \pm 0.09 \pm 0.1 \hphantom{0}$  & ACT, Ref.\cite{ACT:2020frw,Murphy:2024fna} & EB-nulling\tablefootnote{In this case, no reduction of foreground emissions was applied.\cite{ACT:2020frw}.} + optical modeling\\
$\beta$ & $-0.35 \pm 0.14\hphantom{0}$  & Planck (PR3), Ref.\cite{Minami:2020odp} & MK \\
$\beta$ & $-0.30 \pm 0.11\hphantom{0}$  & Planck (PR4), Ref.\cite{Diego-Palazuelos:2022dsq} & MK \\
$\beta$ & $-0.35 \pm 0.70\hphantom{0}$  & Planck LFI + WMAP, Ref.\cite{Cosmoglobe:2023pgf} & MK \\
$\beta$ & $-0.34 \pm 0.09\hphantom{0}$  & Planck + WMAP, Ref.\cite{Eskilt:2022cff} & MK \\
$\beta$ & $-0.37 \pm 0.11\hphantom{0}$  & Planck (HFI + LFI), Ref.\cite{Eskilt:2022wav} & MK \\
$\beta$ & $-0.34 \pm 0.09\hphantom{0}$  & Planck + WMAP, Ref.\cite{Eskilt:2022cff} & MK \\
 \botrule
\end{tabular} \label{table1}}
\end{table}

To address the issue of accurately determining $\alpha$ and $\beta$, a new method was proposed in Ref.~\refcite{Minami:2019ruj} and further developed in Ref.~\refcite{Minami:2020fin}. In these papers, the degeneracy between $\alpha$ and $\beta$ is resolved by two key elements: 1) the use of foreground emissions; and 2) the assumption that the CPR/CB effect is proportional to the distance traveled by the photons (as it is in the CS theory). Specifically, photons from the early last scattering surface experience a rotation of the polarization plane given by $\alpha + \beta$, while photons from our Galaxy (foreground emissions) are rotated only by $\alpha$. Applying this idea to all frequencies in the Planck release 3 (PR3), the authors of Ref.~\refcite{Minami:2020odp} obtained $\beta= 0.35 \pm 0.14 \, \mbox{[deg]}$, excluding $\beta=0$ at $99.2\%$ C.L.. A similar analysis with Planck release 4 (PR4) provided $\beta=0.30 \pm 0.11 \, \mbox{[deg]}$ (see Ref.~\refcite{Diego-Palazuelos:2022dsq}) although the authors chose not to assign cosmological significance to this estimate due to remaining uncertainties in foreground polarization.
See also Ref.~\refcite{Diego-Palazuelos:2022cnh} for a comprehensive study of the robustness of this analysis, including the use of templates to model the intrinsic correlation of the $E$ and $B$ modes created by Galactic dust emission.
Interestingly, when only lower frequencies (e.g., WMAP and Planck LFI data) are considered, where Galactic dust emission is expected to be less dominant, the most likely value for $\beta$ is similarly found to be $\beta = 0.35 \pm 0.70 \, \mbox{[deg]}$, although the larger uncertainty renders it compatible with a null effect \cite{Cosmoglobe:2023pgf}. A generalization of these analyses, with $\beta$ measured at each available frequency, is presented in Ref.~\refcite{Eskilt:2022wav}. This analysis indicates that $\beta$ shows no frequency dependence, consistent with expectations from CS extensions of standard electromagnetism, see again Ref.~\refcite{Carroll:1989vb}. Furthermore, including all available data sets (WMAP, Planck LFI, and Planck HFI), it is possible to exclude $\beta= 0$ at $99.987\%$ C.L., equivalent to a $3.6 \sigma$ significance \cite{Eskilt:2022cff}. All these constraints are summarized in Table \ref{table1} (an update of what presented in Ref.~\refcite{diSeregoAlighieri:2015tgl}).

\section{Recent Constraints on the Anisotropic CPR/CB Effect}
\label{Anisotropic}

Anisotropies in CPR/CB are naturally expected in CS theories, as they are driven by spatial fluctuations of the axion-like field, see Refs.~\refcite{Li:2008tma,Caldwell:2011pu}. 
In general, anisotropic CPR/CB provides additional information beyond the isotropic component (see, for instance, Refs.~\refcite{Greco:2022xwj,Greco:2024oie}). 
For example, its correlation with the CMB T-field has been proposed as a probe to interpret the axion-like field in CS theory as an early dark energy field, introduced to address the well-known Hubble tension\cite{Capparelli:2019rtn}. 
See also Ref.~\refcite{Greco:2022ufo} for an analysis involving the cross-spectra and cross-bispectra of CPR/CB with CMB fields, and Ref.~\refcite{Arcari:2024nhw} for a study on cross-correlation with the spatial distribution of galaxies.

Moreover, anisotropic CPR/CB does not suffer from the degeneracy with instrumental polarization angles discussed above. As a result, many new constraints on the anisotropic CPR/CB effect have been derived in recent literature through CMB observations.
It is customary to provide these constraints in term of the scale invariant amplitude, 
$
A^{\beta \beta} \equiv L (L+1) \, C_L / 2 \pi, 
$
where $C_L$ is the angular power spectrum at multipole $L$ of the CPR/CB angle, i.e. $\langle a_{LM} a_{L'M'}^{\star} \rangle = C_L \, \delta_{LL'} \delta_{MM'}$
with $a_{LM}$ being the coefficients of the spherical harmonics expansion of the fluctuation of the CPR/CB angle, defined as $ \delta \beta(\hat n) \equiv \beta(\hat n)-\beta$.
All current constraints on $A^{\beta \beta}$ are compatible with no detection with an uncertainty of $\sim 0.1 \, [\mbox{deg}^2]$ at $95\%$ C.L. as reported in Ref.~\refcite{Contreras:2017sgi} and in Ref.~\refcite{BICEP2:2017lpa} using Planck (PR2) data and Bicep2-Keck data respectively. Similar constraints on Planck PR3 and PR4 data were reported in Refs.~\refcite{Gruppuso:2020kfy,Bortolami:2022whx,Zagatti:2024jxm}, where constraints on the cross-correlation between the CPR/CB and the CMB T, E, and B fields were also found to be compatible with zero.
The parameter $A^{\beta \beta}$ is constrained to $A^{\beta \beta} < 0.033 \, [\mbox{deg}^2]$ at $95\%$ C.L. by ACTPol\cite{Namikawa:2020ffr} and SPTPol\cite{SPT:2020cxx}. 
In the latter reference also the cross-correlation between $\beta$ and the CMB T field was determined  well compatible with null effect.
The current best constraint reads $A^{\beta \beta} < 0.014 \, [\mbox{deg}^2]$ at $95\%$ C.L., obtained from Bicep3-Keck data\cite{BICEPKeck:2022kci}.

\section{Final Remarks}
\label{Conclusions}

The MK technique applied to Planck CMB polarized observations has provided a hint of detection for an isotropic CPR/CB angle at about the $3 \sigma$ confidence level, while the anisotropic component of this effect, as measured through Planck, ACTPol, SPTPol, and Bicep3-Keck data, is currently found to be well compatible with zero.
If these observations are confirmed \cite{Komatsu:2022nvu}, they would suggest parity-violating extensions of standard electromagnetism and, consequently, point to the existence of a new scalar cosmological field potentially acting as dark matter and/or dark energy \cite{Marsh:2015xka,Ferreira:2020fam}. See also e.g. Refs.~\refcite{Fujita:2020ecn,Murai:2022zur,Nakatsuka:2022epj,Greco:2024oie}.

To confirm these signatures, we need: 1) CMB data beyond that from Planck to perform the isotropic CPR/CB analysis; and 2) an approach that avoids using foreground emissions \cite{Clark:2021kze,Hervias-Caimapo:2024ili} to break the degeneracy between $\beta$ and $\alpha$. The latter requirement calls for very accurate calibration of the instrumental polarization angle.
For instance, the Bicep3 collaboration has shown that the BICEP3 2-year dataset (2017–2018) has an on-sky sensitivity to $\beta$ of 0.078 deg, which could be improved to 0.055 deg by including all existing BICEP3 data through 2023 \cite{BICEPKeck:2024cmk}. 
See also Ref.~\refcite{Ritacco:2024kug} for details on the COSMOCal project, which aims to deploy a polarized source in space to calibrate microwave frequency observations.

Future CMB experiments, such as SO, LiteBIRD, and CMBS4, are expected to improve current constraints on $\beta$ and $A^{\beta \beta}$ by one or more orders of magnitude\cite{Pogosian:2019jbt,LiteBIRD:2022cnt}, providing independent and essential measurements of this phenomenon. Upcoming releases of current CMB observations are also highly anticipated. In all cases, the main experimental challenge remains the calibration of the instrumental polarization angle, a critical obstacle to accurately measuring the isotropic CPR/CB angle without relying on foreground emissions, as required by the MK technique.

\section*{Acknowledgments}
We thank Nicola Bartolo, Fabio Finelli, Matteo Galaverni, Alessandro Greco, and Paolo Natoli for valuable discussions. AG acknowledges support from the MUR PRIN2022 Project ‘BROWSEPOL: Beyond the Standard Model with Cosmic Microwave Background Polarization’ (2022EJNZ53), funded by the European Union - Next Generation EU.

%


\end{document}